\documentclass{article}

\usepackage{epsfig}
\usepackage{latexsym}

\begin{document}

\newcommand{\old}[1]{}
\newcommand{\pf}[1]{{\bf Proof:} #1}
\newtheorem{thm}{Theorem}[section]
\newtheorem{lem}[thm]{Lemma}
\newtheorem{cor}[thm]{Corollary}
\newtheorem{rem}[thm]{Remark}
\newtheorem{conjecture}[thm]{Conjecture}
\newtheorem{exam}[thm]{Example}
\newtheorem{claim}[thm]{Claim}
\newtheorem{definition}[thm]{Definition}
\newtheorem{observation}[thm]{Observation}
\newtheorem{fact}[thm]{Fact}

\title{An Algorithmic Study of Manufacturing Paperclips 
       and Other Folded Structures}

\author{Esther M. Arkin\thanks{
Department of Applied Mathematics and Statistics,
  State University of New York, Stony Brook, NY 11794-3600, USA.
Email: [estie,jsbm]@ams.sunysb.edu}
\and
S\'andor P. Fekete\thanks{
{Department of Mathematical Optimization, TU
  Braunschweig, Pockelsstr.~14, D-38106 Braunschweig, Germany.
Email: sandor.fekete@tu-bs.de}}
\and
Joseph S. B. Mitchell\footnotemark[1]
}
\date{}

\maketitle

\begin{abstract}
  We study algorithmic aspects of bending wires and sheet metal into a
  specified structure. Problems of this type are closely related to
  the question of deciding whether a simple non-self-intersecting wire
  structure (a carpenter's ruler) can be straightened, a problem that
  was open for several years and has only recently been solved in the
  affirmative.
  
  If we impose some of the constraints that are imposed by the
  manufacturing process, we obtain quite different results.  
  In particular, we study the variant of the carpenter's ruler problem
  in which there is a restriction that only one joint can be modified
  at a time.  For a linkage that does not self-intersect or
  self-touch, the recent results of Connelly et al. and Streinu imply
  that it can always be straightened, modifying one joint at a time.
  However, we show that for a linkage with even a single vertex
  degeneracy, it becomes NP-hard to decide if it can be straightened
  while altering only one joint at a time.  If we add the restriction
  that each joint can be altered at most once, we show that the
  problem is NP-complete even without vertex degeneracies.

In the special case, arising in wire forming manufacturing, that each
joint can be altered at most once, and must be done sequentially from
one or both ends of the linkage, we give an efficient algorithm to
determine if a linkage can be straightened.

\end{abstract}


\bibliographystyle{abbrv}

\newpage
\section{Introduction}
The following is an algorithmic problem that arises in the study of the
manufacturability of sheet metal parts: {\em Given a flat piece, $F$,
of sheet metal (or cardboard, or other bendable stiff sheet material), can
a desired final polyhedral part, $P$, be made from it?}   The
2-dimensional version is the wire-bending 
(``paperclip'') problem: {\em Given a straight piece, $F$, of wire, can
a desired simple polygonal chain, $P$, be made from it?}  This
problem also arises in the fabrication of hydraulic
tubes, e.g., in airplane manufacturing.\footnote{We thank Karel Zikan
for introducing to us the hydraulic tube bending problem at Boeing's factory.}
In both versions of the problem, we require that any intermediate
configuration during the manufacture of the part be feasible, meaning
that it
should not be self-intersecting.  In particular, the paperclips
that we manufacture are not allowed to be ``pretzels'' -- we assume
that the wire must stay within the plane, and not cross over
itself. See Figure~\ref{fig:paperclips} for an illustration.
We acknowledge that some real paperclips are designed
to cross over themselves, such as the butterfly style of clip shown
in the figure.

Our problem is one of automated process planning: Determine a sequence
(if one exists) for performing the bend operations in sheet metal
manufacturing.  We take a somewhat idealized approach in this paper,
in that we do not attempt to model here the important aspects of tool
setup, grasp positions, robot motion plans, or specific sheet metal
material properties which may affect the process.  Instead, we focus
on the precise algorithmic problem of determining a sequence for bend
operations, on a given sheet of material with given bend lines,
assuming that the only constraint to performing a bend along a given
bend line is whether or not the structure intersects itself at any
time during the bend operation.  

Note that the problem of determining if a bend sequence exists that
allows a structure to {\em unfold} is equivalent to that of
determining if a bend sequence exists that allows one to {\em fold} a
flat (or straight) input into the desired final structure: the bending
operations can simply be reversed.  For the remainder of the paper, we
will speak only of unfolding or straightening.

\subsection{Motivation and Related Work}  

Our foldability problem is motivated from process planning in
manufacturing of structures from wire, tubing, sheet metal, and
cardboard.  
The CAD/CAM scientific community has studied extensively the problem of
manufacturability of sheet metal structures; see the thesis of
Wang~\cite{w-mddsmp-97} for a survey.  Systems have been built (e.g.,
PART-S~\cite{vvsk-parts-92} and BendCad~\cite{gbkk-appsmbo-98}) to do
computer-aided process planning in the context of sheet metal manufacturing; see also
\cite{Ayyadevara_1999_3051,Bourne_1995_2029,Kim_1998_1134,Wang_1997_2031,Wang_1996_2028}.
See \cite{la-fcf:mpa-99}
for a motion planning approach to the problem of
computing folding sequences for folding three-dimensional cardboard cartons.
Considerable effort has gone into the design of good heuristics for
determining a bend sequence; however, the known algorithms are based
on heuristic search (e.g., A$^*$) in large state spaces; they are known to be
worst-case exponential.  (Wang~\cite{w-mddsmp-97} cites the known
complexity as $O(n! 2^n)$.)

Our work is also motivated by the mathematical study of origami, which
has received considerable attention in recent years.  In mathematics
of origami, Bern and Hayes \cite{bh-cfo-96} have studied the
algorithmic complexity of deciding if a given crease pattern can be
folded flat; they give an NP-hardness proof.
Lang~\cite{l-maod-94,l-caod-96} gives algorithms for computing crease
patterns in order to achieve desired shapes in three dimensions.
Other work on computational origami includes
\cite{MapFoldingWADS2001,ddl-foscs-99,ddm-ffswp-00,h-mfo-94,o-cgc33-98,ORo-JCDCG-00,so-cru-87}.
A closely related problem is that of flat foldings of polyhedra.  It
is a classic open question whether or not every convex polytope in
three dimensions can be cut open along its edges so that it unfolds
flat, without overlaps.  Other variants and special cases have been
studied; see
\cite{ao-nsu-92,Ununfoldable,bddloorw-uscop-98,lo-wcpfp-96,nf-u3dcp-93}.

Finally, we are motivated by the study of linkage problems; in fact,
in the time since this paper was first drafted, the carpenter's ruler
conjecture has been resolved by Connelly, Demaine, and
Rote~\cite{cdr-epcbu-00} and Streinu~\cite{s-capnr-00}: Any (strongly)
simple polygonal {\em linkage} with fixed length links and hinged
joints, can be straightened while maintaining strong simplicity (i.e.,
without the linkage crossing or touching itself). (They also show
related facts about linkage systems, e.g., that any simple polygonal
linkage can be {\em convexified}.) In fact, Streinu~\cite{s-capnr-00}
gives an algorithmic solution that bounds the complexity of the
unfolding and is somewhat more general than the slightly earlier
results of \cite{cdr-epcbu-00}.  These results imply that any
(strongly simple) paperclip can be manufactured if one has a machine
that can perform a sufficiently rich set of bending operations.
For a recent overview of folding and unfolding, see the thesis of
Demaine~\cite{dthesis}. 
Earlier and related work on linkages includes
\cite{3DChains_DCG2001,LockedTreeDAM,lw-tpio-91,lw-rcpce-95,pw-rrc-96,pw-frrp-97,ksw-frit-96}.
Our hardness results are particularly interesting and relevant in
light of these new developments, since we show that even slight
changes in the assumptions about the model or the allowed input
results in linkages that cannot be straightened, and it is NP-hard to
decide if they can be straightened.
  
\subsection{Summary of Results}

\begin{description}
  
\item[(1)] We show that it is (weakly) NP-complete to determine if a
  given rectilinear polygonal linkage can be straightened, under the
  restriction that only one joint at a time is altered and each joint
  can be altered only once (so the joint must be straightened in a
  single bend operation).  A consequence is that the more general sheet metal
  bending problem is hard as well, even in the case of parallel bend
  lines and an orthohedral structure~$P$.
  
\item[(2)] We prove that it is (weakly) NP-hard to determine if a
  given polygonal linkage can be straightened if there is a {\em vertex 
  degeneracy}, in which two vertices coincide.  Here we again assume that only
one joint can be altered at a time, but we do not assume 
that a joint is altered only once, so we may make any number of bends
at any particular joint.

\item[(3)] We give efficient algorithms for determining if a given
  bend sequence is feasible, assuming only one joint is altered at a
  time, and for determining if certain special classes of bend
  sequences are feasible.  In particular, we give an efficient
  ($O(n\log^2 n)$) algorithm for determining if a polygonal linkage
  can be straightened using a {\em sequential} strategy, in which the
  joints are completely straightened, one by one, in order along the
  linkage.  We also give efficient polynomial-time algorithms for
  deciding whether there is a feasible bend sequence that straightens
  joints in an order ``inwards'' from both ends or ``outwards''
  towards both ends.  (Such constrained bend sequences may be required
  for automated wire-bending machines.)  These results will be made
more precise in Section~\ref{sec:algorithms}.

\end{description}

\section{Preliminaries}

The input to our problem is a simple polygonal chain (linkage), $P$,
with vertex sequence $(b_0,b_1,b_2,\ldots,b_{n+1})$.  The points $b_0$
and $b_{n+1}$ are the {\em endpoints} of the chain, and the $n$
vertices $b_1,\ldots,b_n$ are the {\em bends} (or {\em joints}).  The
line segments $b_ib_{i+1}$ are the {\em edges} (or {\em links}) of
$P$.  The edge $b_ib_{i+1}$ is a closed line segment; i.e., it includes its
endpoints.  We consider the chain $P$ to be oriented from $b_0$ to
$b_{n+1}$, and we consider each edge of $P$ to have a {\em left} and a
{\em right} side.  Each bend $b_i$ has an associated {\em bend angle}
$\theta_i\in (0,2\pi]$, measured between the right sides of the two
edges incident on $b_i$.  

The chain $P$ is {\em strongly simple} if any two edges, $b_ib_{i+1}$
and $b_jb_{j+1}$, of $P$ that are not adjacent ($i\neq j$) are
disjoint and any two adjacent edges share only their one common
endpoint.  We say that $P$ is {\em simple} if it is not self-crossing
but it possibly is self-touching, with a joint falling exactly on a
non-incident edge or another joint; i.e., $P$ is simple if it
is strongly simple or an infinitesimal perturbation of it is strongly
simple.

We consider the chain $P$ to be a structure consisting of rigid rods
as edges, whose lengths cannot change, connected by hinged joints.
When a {\em bend operation} is performed at joint $b_i$, the bend
angle $\theta_i$ is changed.  Throughout this paper, we assume that
the only bend operations allowed are {\em single-joint} bends, in
which only one bend angle is altered at a time.  We establish the
convention that when a bend operation occurs at $b_i$, the subchain
containing the endpoint $b_0$ remains fixed in the plane, while the
subchain containing $b_{n+1}$ rotates about the joint $b_i$.  This
convention allows us to have a unique embedding of a partially or
fully straightened chain in the plane.

A bend operation is {\em complete} if, at the end of the operation,
the bend angle is $\pi$; we then say that the joint has been {\em
  straightened}.  A bend operation that is not complete is called a
{\em partial bend}. A sequence of bend operations is said to be 
{\em monotonic} if no bend operation 
increases the absolute deviation from straightness, $|\theta_i-\pi|$,
for a joint $b_i$.
If all joints of $P$ have been straightened, the
resulting chain is a straight line segment, $F$, of length
$\sum_{i=0}^n |b_ib_{i+1}|$, where $|b_ib_{i+1}|$ denotes the
Euclidean length of segment $b_ib_{i+1}$.  By our bend operation
convention, one endpoint of $F$ is $b_0$, and $F$ contains the segment
$b_0b_1$ (which never moves during bend operations).

For $S\subseteq B$, we let $P(S)$, denote the partially straightened
polygonal chain having each of the bends $b_i\in S$ straightened (to
bend angle $\pi$), while each of the other bends, $b_i\notin S$, is at
its original bend angle $\theta_i$.  Thus, in this notation $P(B)=F$
and $P(\emptyset)=P$.  We let $P(S;i,\theta)$, for $1\leq i\leq n$
with $i\notin S$, denote the chain in which each bend $b_j\in S$ is at
bend angle $\pi$, bend $b_i$ is at angle $\theta$, and all other bends
$b_j\notin S$ are at their original bend angles $\theta_j$.  We say
that chain $P(S)$ or $P(S;i,\theta)$ is {\em feasible} if it is a
simple chain.

We say that bend $b_i$ is {\em foldable} (or is a {\em feasible fold})
for $P(S)$ if $P(S;i,\theta)$ is feasible for all $\theta$ in the
range between $\pi$ and $\theta_i$ (more precisely, for all
$\theta_i\leq \theta\leq \pi$, if $\theta_i<\pi$, or for all $\pi\leq
\theta\leq \theta_i$, if $\theta_i>\pi$).  If $b_i$ is foldable, then
it is possible to make a complete bend at $b_i$, meaning that the
joint can be straightened in a single operation without causing the
chain to self-intersect.  We say that a permutation
$\sigma=(i_1,i_2,\ldots,i_n)$ of the indices $\{1,2,\ldots,n\}$ is
{\em foldable for $P$} if, for $j=1,2,\ldots,n$, joint $b_{i_j}$ is
foldable for $P(\{b_{i_1},\ldots,b_{i_{j-1}}\})$, i.e., if $P$ can be
unfolded into the straight segment $F$ using the bend sequence
$\sigma$ (so that, by reversing the operations, $P$ can be
manufactured from $F$ using the reverse of the bend sequence).

The {\sc Wire Bend Sequencing} problem can be formally stated as: {\em
  Determine a foldable permutation $\sigma$, if one exists, for a
  given chain~$P$.}

This paper studies the {\sc Wire Bend Sequencing} problem for
polygonal chains in the plane.  We note, however, that our results
have some immediate implications for the {\sc Sheet Metal Bend
  Sequencing} problem, which is defined analogously for a polyhedral
surface $P$ having a pattern $B$ of bend lines (creases), each of
which must be straightened in order to flatten $P$ into a flat polygon
$F$.  Specifically, the hardness of the {\sc Sheet Metal Bend
  Sequencing} follows from the hardness of the {\sc Wire Bend
  Sequencing}, which can be seen as a special case of the sheet metal
problem in which $F$ is a rectangle and the bend lines $B$ are all
segments parallel to two of the sides of $F$ and extending all the way
across~$F$.

\begin{figure}[htbp]
\begin{center}
  \leavevmode
  \epsfig{file=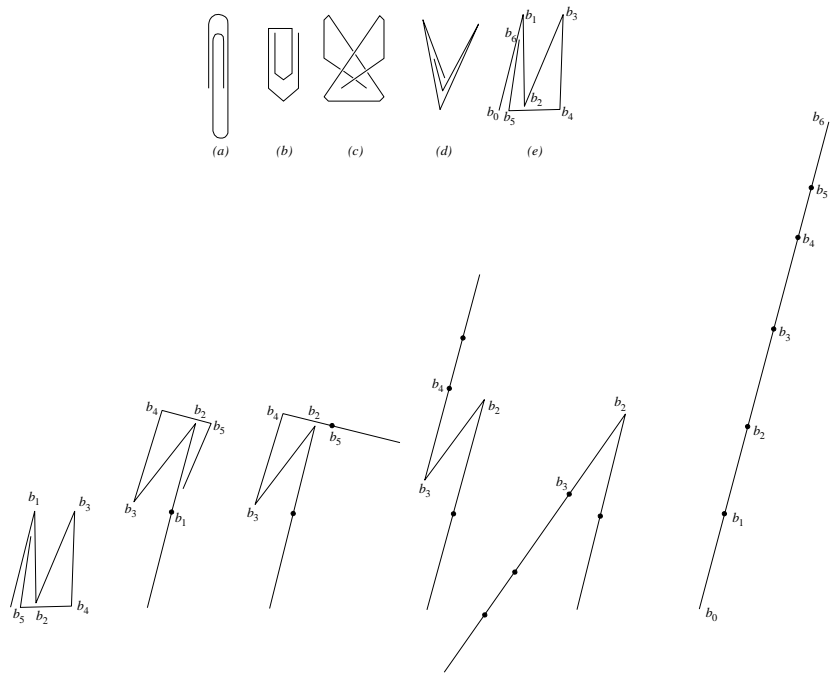,width=0.9\textwidth}
  \caption{Examples of paperclips: (a) and (b) are standard versions, which are
readily straightened. (c) is a ``butterfly'' paperclip, which is not a planar structure and is not among the wire structures considered in our two-dimensional model.  (d) shows a 5-link paperclip that cannot be straightened using
complete bends in the plane. (e) shows a 6-link structure that can be
straightened, e.g., using the bend sequence animated below it for the
bend sequence $\sigma=(1,5,4,3,2)$.}
  \label{fig:paperclips}
\end{center}
\end{figure}

We give an example in Figure~\ref{fig:paperclips} of some common
paperclip shapes, (a)--(c).  We also show an example, (d), of a 5-link
paperclip that cannot be straightened using complete bends, for any
permutation $\sigma$ of the bends.  Finally, we show an example of a
6-link paperclip for which the foldable permutations are
$\{(1,5,4,3,2), (1,5,4,2,3)\}$; we show the sequence of bends, with
the intermediate structures, for the permutation $\sigma=(1,5,4,3,2)$.

\vspace*{20mm}
\clearpage

\section{Hardness Results}

\subsection{Complete Bends}

Our first result shows that if we require bends to be complete, as in
our specification of the {\sc Wire Bend Sequencing} problem, the
problem of deciding if there is a feasible bend sequence is
NP-complete.  

\begin{thm}
\label{th:complete}
{\sc Wire Bend Sequencing} is (weakly) NP-complete, even
if $P$ is rectilinear.
\end{thm}

\begin{pf}
  We prove NP-completeness, even in the case that we are restricted to
  a special class of bend sequences, namely, those that can be written
  as the concatenation of up to four monotone subsequences of the
  index set $\{1,\ldots,n\}$.  Below, we refer to each subsequence of
  bends as a {\em monotone pass} over the chain, going from one end to
  the other, performing a specified subset of complete bends.

Our reduction is from {\sc Partition}:  Given a set $S$ of 
$n$ integers, $a_i$, which sum to $A=\sum_i a_i$,
determine if there exists a partition of the set into two 
subsets each of which sums to~$A/2$.

\begin{figure}[htbp]
\begin{center}
  \leavevmode
  \epsfig{file=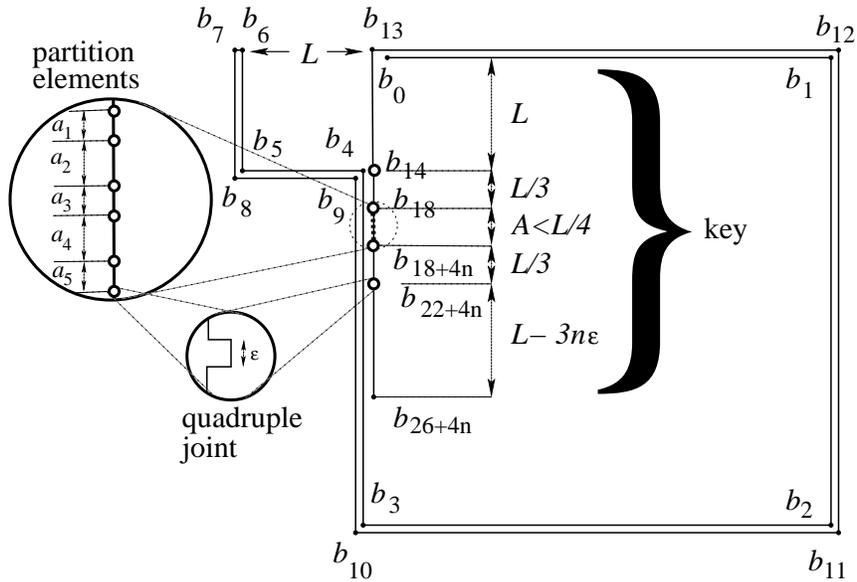,width=0.95\columnwidth}   
  \caption{Proving hardness of the {\sc Wire Bend Sequencing} problem for rectilinear
  chains: Frame and key.}
  \label{fi:frame}
\end{center}
\end{figure}

The key idea of our construction uses two components,
as shown in Figure~\ref{fi:frame}:
One is a rigid ``frame'' that can only be unfolded
if one end of the chain can be removed from within this frame.
The other component is a ``key'' that encodes the partition instance.
Collapsing the key is possible if and only if there is a partition
of the integers into two sets of equal sum. The total number
of segments will be $\ell=26+4n$; we write $b_i$ ($i=0,1,\ldots,26+4n$) for the
vertices, and $s_i=(b_{i-1},b_i)$ for the segments.
For any point in time, we refer to the position of a joint $b_i$
by its coordinates $(x_i,y_i)$. When discussing some of the relative distances,
we use $d_{\infty}(b_i,b_j)=\max\{|x_i-x_j|,|y_i-y_j|\}$.

More precisely, the frame
consist of $13$ segments, $s_1=(b_0,b_1),\ldots,s_{13}=(b_{12},b_{13})$,
as shown in the figure. Segment lengths are chosen such that
the size of the frame is $\Theta(L)$,
with minimal coordinate differences 
$d_{\infty}(b_0,b_{13})$,
$d_{\infty}(b_1,b_{12})$,
$d_{\infty}(b_2,b_{11})$,
$d_{\infty}(b_3,b_{10})$,
$d_{\infty}(b_4,b_{9})$,
$d_{\infty}(b_5,b_{8})$,
$d_{\infty}(b_6,b_{7})$
being $\Theta(\varepsilon)$,
where $\varepsilon=1/(n^3L^2)$.
The ``key'' consists of $13+4n$ segments, 
$s_{14}=(b_{13},b_{14})$, $\ldots, s_{26+4n}=(b_{25+4n},b_{26+4n})$.
For $i=0,\ldots,4n+2$, the ``auxiliary'' segments $s_{15+4i}$,
$s_{16+4i}$, 
$s_{17+4i}$ have length $\varepsilon$, 
while the ``partition'' segments $s_{18+4i}$
have length $a_i$. The long ``positioning'' segments $s_{14}$,
$s_{15}$,
$s_{25+4n}$,
$s_{26+4n}$
have lengths $L$, $L/3$, $L/3$, and $L-3n\varepsilon$, respectively;
they guarantee that the partition segments must have a particular
relative position when removing the key. 
We choose the scale to be such that
$L/4>A$, for technical reasons that will become clear
later in the proof.
As indicated in the figure,
the initial position of each key segment $s_i, i=14,\ldots,26+4n$
has $x$-coordinate $x_{13}$ or $x_{13}+\varepsilon$, 
with a horizontal distance of 
$x_i-x_4=\varepsilon$ or
$x_i-x_4=2\varepsilon$ 
from
$s_{4}$. Moreover, $b_{14}$ is positioned at a vertical
distance of $y_{14}-y_4=n^2\varepsilon=1/(nL^2)$ above $b_4$.

The purpose of the auxiliary segments is as follows.
As shown in Figure~\ref{fi:frame}, we have two types of joints in the figure:
the ``ordinary'' ones (indicated by solid black dots)
form the frame and can only be accessed once.
The ``quadruple'' ones (indicated by hollow dots in 
Figure~\ref{fi:frame}) consist of the four simple joints at three
consecutive auxiliary segments; they are found along the key as described.
These quadruple joints make 
it possible to simulate opening and closing 
such a joint a limited number of times.

\begin{figure}[htbp]
\begin{center}
  \leavevmode
  \epsfig{file=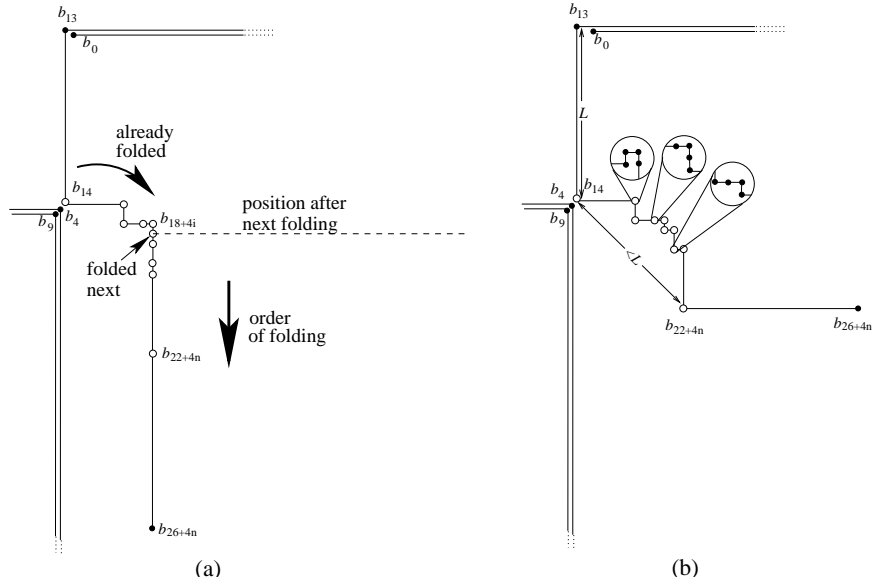,width=0.95\columnwidth}  
  \caption{Turning the key into a stair: (a) An intermediate stage
  of the monotone pass. (b) The stair configuration at the end of
  the monotone pass, with details of the state of quadruple joints.}
  \label{fi:stair}
\end{center}
\end{figure}

Now assume that there is a partition $S=S_1\stackrel{\cdot}{\cup}S_2$, 
such that $\sum_{i\in S_1} a_i=\sum_{i\in S_2} a_i$. 
In order to see that the key can be removed from the frame
we first convert it
into the ``stair'' configuration shown in Figure~\ref{fi:stair}:
We make one monotone pass over the chain towards the key end,
and straighten one ordinary joint per quadruple joint
whenever this joint separates two segments from different
$S_i$. Thus, segments corresponding to numbers in 
$S_1$ will be horizontal, while those
for numbers in $S_2$ will be vertical.
In order to keep the number of monotone passes limited to four,
during this first pass
we also straighten two ordinary joints per quadruple joint 
separating two segments from the same
$S_i$, as shown in Figure~\ref{fi:stair}(b).

\begin{figure}[htbp]
\begin{center}
  \leavevmode
  \epsfig{file=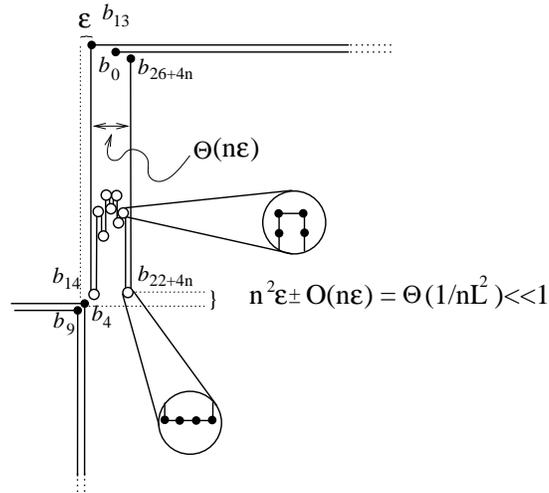,width=0.6\columnwidth}  
  \caption{Turning the stair into a harmonica of small width
and length $L$. (The horizontal width is not drawn to scale
in order to show details.)}
  \label{fi:harmony}
\end{center}
\end{figure}

Making a similar monotone pass, we
can convert the stair into a ``flat harmonica'', as shown
in Figure~\ref{fi:harmony}, with 
segments from $S_2$ pointing ``down'', i.e., $y_i<y_{i-1}$, 
and segments from
$S_1$ pointing ``up'', i.e., $y_i>y_{i-1}$. By assumption about the 
partition, the positions of 
endpoints $b_{18}$ and $b_{18+4n}$ satisfy
$d_{\infty}(b_{18},b_{18+4n})<3n\varepsilon$, 
and $2L/3-3\varepsilon<y_{13}-y_{18}<2L/3+3\varepsilon$,
i.e., both $b_{18}$ and $b_{18+4n}$ are roughly $2L/3$ below $b_{13}$.
Altogether, the position of the last segment $s_{26+4n}$
of length $L$ in the chain will differ by at most $O(n\varepsilon)$
from the
vertical position of segment $s_{14}$, with all 
other segments strictly in-between. This collapsed structure
can be rotated about $b_{13}$ without colliding with any frame
segments. Then it is easy to open up the remaining frame
(by straightening $b_{12}$, $b_{11}$, $b_{10}$, $b_8$, 
$b_7$, $b_6$, $b_5$, $b_4$, $b_3$, $b_2$, $b_1$ as one monotone pass, skipping
$b_9$.)
Finally, the resulting monotone chain can be straightened
in one last monotone pass. 

Conversely, assume now that the chain can be straightened.
See Figure~\ref{fi:circle}.
It is clear that $b_{13}$ must be straightened before
any other joint in the set $\{b_1,\ldots,b_{12}\}$.
In order to avoid hitting vertex $b_4$ 
during this motion, any part of the key to the right and below
$b_{13}$ must be strictly within the circle $C$ of radius 
$r=\sqrt{(L+1/(nL^2))^2+\varepsilon^2}<L+2/(nL^2)$ around
$b_{13}$, where $r$ is the distance between $b_{13}$ and~$b_4$ 
(see Figure~\ref{fi:harmony}.) The following technical arguments
show that at this time,
segment $s_{26+4n}$ has to be in a vertical position that basically
coincides with $s_{14}$, which is only possible in case of
a feasible partition.

\begin{figure}[h!tbp]
\begin{center}
  \leavevmode
  \epsfig{file=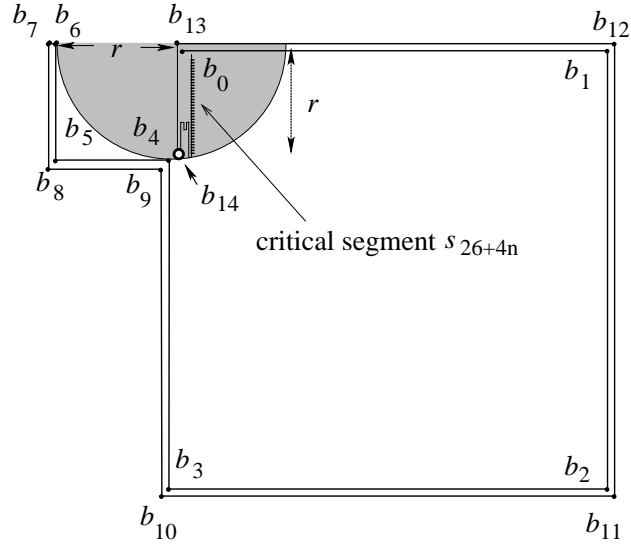,width=0.75\columnwidth}  
  \caption{When straightening joint $b_{13}$, the key must be
  fully contained in the shaded circle of radius $r<L+2/(nL^2)$. This forces
  a particular position of segment~$s_{26+4n}$.}
  \label{fi:circle}
\end{center}
\end{figure}

When starting the rotation about $b_{13}$, 
$s_{26+4n}$ is an axis-parallel
segment of length $L-3n\varepsilon>L-1/(nL^2)$.
The rigid frame and the closeness of $b_{14}$ and $b_4$ ensure that segment
$s_{26+4n}$ cannot lie to the left of $s_{14}$, implying
that $s_{26+4n}$ can only lie within the quarter circle 
of radius $r$ below and to the right of $b_{13}$
when $b_{13}$ is straightened. 

Let $b_{\min}$ be one of the two points in 
$\{b_{25+4n},b_{26+4n}\}$ that is not further from $b_{13}$ than the other, 
and let $b_{\max}$ be the other point.
If the vertical distance
$y_{13}-y_{\min}$ is greater than $\sqrt{7/(nL)}$, it follows
that the Euclidean distance between $b_{\max}$ and $b_{13}$ is at least
$$\sqrt{\frac{7}{nL}+\left(L-\frac{1}{nL^2}\right)^2}
=\sqrt{\left(L+\frac{2}{nL^2}\right)^2+\frac{1}{nL}-\frac{3}{n^2L^4}}
>L+\frac{2}{nL^2} >r,$$
a contradiction
to the assumption that $s_{26+4n}$ is fully contained in $C$.
Now, using our assumption that $L/4>A$, we know that
$b_{25+4n}$ and $b_{14}$ are connected by a polygonal chain of
length strictly less than
$L/3+L/4+L/3=11L/12$, implying that $b_{25+4n}$ has Euclidean distance
at least $L/12$ from $b_{13}$, so $b_{26+4n}=b_{\min}$ and
$b_{25+4n}=b_{\max}$. As $b_{26+4n}$ is within $\sqrt{7/(nL)}$
of $b_{13}$, it follows that $b_{26+4n}$ has Euclidean
distance at least
$L-\sqrt{7/(nL)}$ from $b_{14}$. If $s_{26+4n}$ were horizontal, then
the Euclidean distance between 
$b_{25+4n}$ and $b_{14}$ would be at least 
$\sqrt{\left(L-\sqrt{7/(nL)}\right)^2+(L-1/(nL^2))^2}> 11L/12$, a contradiction. 
Hence, $s_{26+4n}$ must be vertical.
Just as we derived for the vertical distance between
$b_{13}$ and $b_{26+4n}$, it follows for
the horizontal distance that $x_{26+4n}-x_{13}\leq \sqrt{7/(nL)}\ll 1$.

Now observe that when starting the rotation about
$b_{13}$, all partition segments must be strictly between $s_{14}$ and
the narrow strip between $s_{14}$ 
and $s_{26+4n}$, meaning that they are all vertical. 
Let $S_1$ be the set of ``upwards'' partition segments $s_i$ with
$y_i>y_{i-1}$, and $S_2$ be the set of ``downwards''
partition segments $s_i$ with $y_i<y_{i-1}$.
As 
$|y_{24+4n}-y_{15}|=\Theta(n\varepsilon)$ and
$|y_{25+4n}-y_{14}| =\Theta(n^2\varepsilon)$, we conclude that
the integral total length of upwards segments equals the integral
total length of downwards segments. 

This means 
that $\sum_{i\in S_1}a_i=\sum_{i\in S_2}a_i$, and we have a feasible partition.
This completes the proof. 
\hfill $\Box$
\end{pf}

\subsection{Partial Bends}

Now we consider the case in which each joint may be changed an
arbitrary number of times during the straightening operations, while
still making single-joint bends (bending only one joint at a time).
This version of the problem is closely related to the carpenter's
ruler problem studied by~\cite{cdr-epcbu-00,s-capnr-00}.  
In the context of our study on folding, there may be the additional
requirement of using only monotonic bend operations,
e.g., to avoid work-hardening the wire, possibly causing it to break.
We begin with the following observation about the sufficiency
of monotonic single-joint bends;
see also the discussion on p.~9 of Demaine's 
thesis~\cite{dthesis}.

\begin{thm}
\label{th:simple}
Any strongly simple polygonal chain $P$ can be straightened using
a finite number of monotonic single-joint bends.
\end{thm}

\begin{pf}
  Consider the set $\mathcal{S}$ of points in $n$-dimensional
  joint-angle space that correspond to strongly simple embeddings of
  the linkage.  A single-joint bend corresponds to axis-parallel
  motion in joint-angle space.  If self-touching is prohibited,
  $\mathcal{S}$ is an open set; note too that $\mathcal{S}$ is
  bounded.  By Streinu's result~\cite{s-capnr-00}, there is an opening
  motion of the chain that consists of a finite number of individual
  monotonic moves.  Such an opening motion corresponds to a path,
  $\Pi$, in $\mathcal{S}$, comprised of a finite number of arcs, each
  corresponding to a monotonic move.  Let $\varepsilon$ be the
  Euclidean distance between path $\Pi$ and the boundary of
  $\mathcal{S}$; since $\mathcal{S}$ is open, we know that
  $\varepsilon>0$.  Then we can replace each arc of the path $\Pi$
  with a finite sequence of axis-parallel moves of size
  $\varepsilon/2$, yielding a straightening that uses single-joint
  bends.  \hfill $\Box$
\end{pf}

We will refer to a sequence of small individual moves that mimics an
overall large-scale motion of several joints as ``wiggly'', since the
overall motion may be achieved through back-and-forth motions of
individual segments that gradually change individual angles.

The following results show that allowing even a single
point of self-incidence along the linkage changes the overall
situation quite drastically.

\begin{lem}
\label{le:rigid}
There are polygonal chains $P$ with a single
vertex-to-vertex incidence that cannot be straightened using
partial single-joint bends.
\end{lem}

\begin{figure}[htbp]
\begin{center}
  \leavevmode
  \epsfig{file=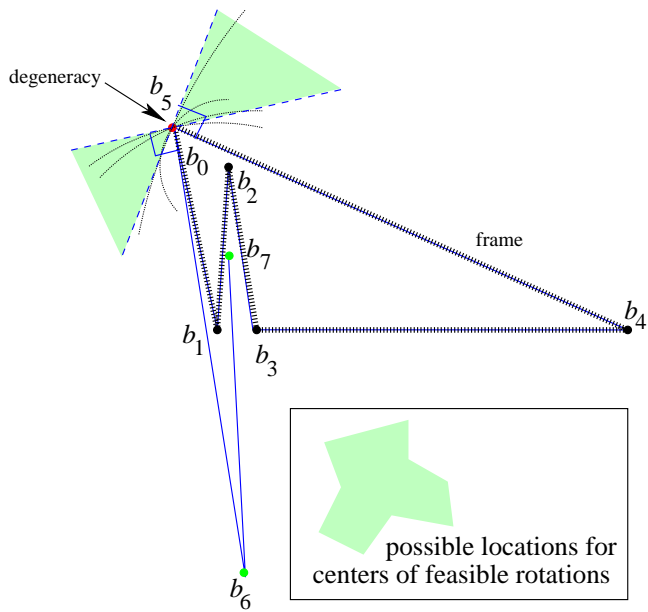,width=0.7\columnwidth}  
  \caption{A polygonal chain that cannot be opened with single-joint
  moves.}
  \label{fi:rigid}
\end{center}
\end{figure}

\begin{pf}
See Figure~\ref{fi:rigid}. The chain has eight joints
(labeled $b_0,\ldots,b_7$) and seven segments (of the form
$s_i=(b_{i-1},b_i)$). The endpoint $b_0$ coincides with joint
$b_5$. It is easily checked that none of the joints
$b_1,\ldots,b_4$ can be changed without causing a self-intersection:
Assume that there is a feasible
motion of a joint $b_i$ with $0<i<{5}$. Then the points $b_0$
and $b_{5}$ would move away from each other along a circle
around $b_i$. Without loss of generality, assume that 
$b_{5}$ remains in place, while $b_0$ is moving.
Now consider the first such rotation 
that starts with $b_0$ and $b_5$ coinciding, and 
that avoids a crossing of $s_1$ with both
$s_{5}$ and $s_{6}$. If $b_0$ moves clockwise 
around $b_i$, it is easy to see that the angle between
$(b_0,b_i)$ and $s_{5}$ must be at least 
$\pi/2$ when starting the motion, or else $s_1$
and $s_{5}$ intersect. 
If $b_0$ moves counterclockwise around $b_i$, the same
follows for the angle between $(b_0,b_i)$ and
$s_{6}$.  Therefore, the center of rotation must lie
within the shaded region shown in the figure.
(The cone to the left of $b_5$ is feasible for clockwise rotation,
while the cone to the right of $b_5$ is feasible for counterclockwise
roation.)
However, none of the joints
$b_1,\ldots,b_4$ lies inside of this feasible
region. It follows that $b_0,\ldots,b_5$ form a rigid frame,
as long as the angle at $b_5$ stays smaller than $\pi/2$.

On the other hand, it is easy to see that $b_7$ cannot be removed
from the pocket formed by $b_1$, $b_2$, and $b_3$ if only the
two remaining ``free'' joints $b_5$ and $b_6$ can be changed.
The claim follows.
\hfill $\Box$
\end{pf}

If $b_0$ and $b_5$ have some positive distance, then the frame can be opened
along the lines of the approach in \cite{cdr-epcbu-00} or
\cite{s-capnr-00}
by gradually straightening $b_5$, $b_6$, $b_1$, $b_2$, and $b_3$,
so that the ``zig-zagging'' part between $b_0$ and $b_3$ pushes left,
while $b_6$ swings around $b_5$.

Using the frame as a gadget, we can show the following:

\begin{thm}
\label{th:degenerate}
It is NP-hard to decide if a polygonal chain $P$ with a single
vertex-to-vertex incidence can be straightened by 
arbitrary partial single-joint bends.
\end{thm}

\begin{pf}
The basic idea is similar to the one in
Theorem~\ref{th:complete} and also establishes
a reduction of {\sc Partition}. (Refer to Figure~\ref{fi:degenerate}
for an overview.) As before, we write $b_i$ for the joints,
and $s_i=(b_{i-1}, b_i)$ for the segments.
We use the idea of the construction from Lemma~\ref{le:rigid}
to construct a rigid frame, with the key corresponding
to the free end of that chain.
The frame has one end, $b_0$, of the polygonal chain 
wedged into the corner $b_{13}$, 
which has angle $\varphi\ll\pi/2$.
Because of the degeneracy at $b_{13}$, none of the joints $b_1,\ldots,b_{12}$
can be moved individually without causing a self-intersection between
$b_0$ and the chain in the neighborhood of $b_{13}$:
Just like in the proof of Lemma~\ref{le:rigid}, none of the joints 
$b_1,\ldots,b_{12}$ lies in the area of possible locations of feasible
rotations. This continues to be the case while 
$\phi+\psi<\pi$, i.e., while
the sum of angles at $b_{13}$ and at $b_{17}$ does not change significantly.

Again, the ``key'' contains the 
$n$ segments $s_{19},\ldots,s_{19+n}$
of integral lengths $a_1,\ldots, a_n$ that encode 
an instance of {\sc Partition}.
As before, let $S$ denote the set of 
integers for the {\sc Partition} instance. We also use ``long'' auxiliary 
segments of lengths $L/2$ and $L$, where $L\gg \sum_i a_i=A$. Here segments
$s_{19}$ and $s_{20+n}$ have length $L/2$, while $s_{21+n}$ has length $L$. 

The critical dimensions
of the frame are chosen such that the key can just be removed 
from the frame if and only if it can be collapsed to a length of $L$.
Removing the key consists in pulling it through the narrow bottleneck formed
by the segments $s_3$ and $s_{9}$
by extending the ``spring'' formed by $s_{14}$ and $s_{15}$,
while moving the ``keyholder'' $s_{18}$ down 
by a distance of $L+\varepsilon$.
This is possible if and only if there is a feasible partition.

\begin{figure}[htbp]
\begin{center}
  \leavevmode
  \epsfig{file=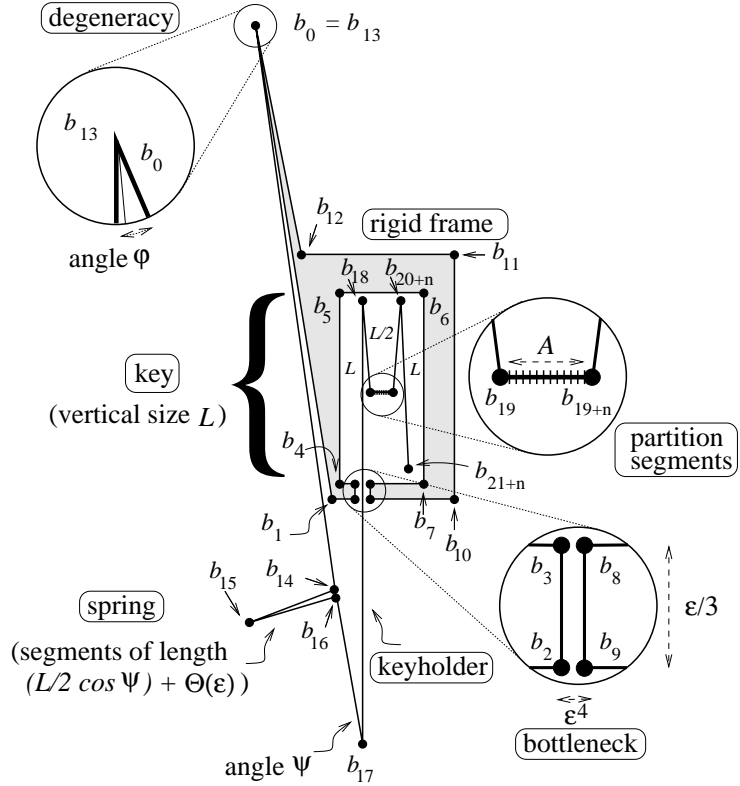,width=0.80\columnwidth}  
  \caption{Illustration of the proof of Theorem~\protect\ref{th:degenerate}.
Note that lengths are not drawn to scale, in order to show sufficient details;
in particular, the dimensions of the bottleneck are much smaller than the
edges encoding the partition instance.}
  \label{fi:degenerate}
\end{center}
\end{figure}

More precisely, let the angle at $b_{17}$ be $\psi\ll\pi$. 
Let $\varepsilon=O(1/nL)$. We assume that the dimensions of the frame
are chosen sufficiently large to guarantee that moving
$b_{17}$ down by a vertical distance of $L+\varepsilon$ 
increases its distance from $b_{13}$
by $L\cos\psi+\Theta(\varepsilon)$, i.e., the 
angles at $b_{17}$ and at $b_{13}$ do not change much. 
The segments
$s_{14}$ and $s_{15}$ forming the spring have length
$L/2\cos\psi+\Theta(\varepsilon)$, so extending the spring
will just suffice to move the keyholder $s_{18}$ down by $L+\varepsilon$,
but not more.
The vertical ``height'' of the bottleneck, i.e., the length of the segments
$s_3$ and $s_9$, is $\varepsilon/3$, while the horizontal ``width''
$x_8-x_3=x_9-x_2=\varepsilon^4$ is significantly smaller.
(As we will discuss below, this forces the keyholder to be roughly vertical
throughout the motion.)
For the initial position of the key inside of the frame, we assume
$L<y_{18}-y_3<L+\varepsilon/3$ and
$L<y_{20+n}-y_3<L+\varepsilon/3$.
Finally, $y_6-y_7=L+\Theta(\varepsilon)$ and $x_6-x_5=\Theta(A)$, so
the dimensions of the rectangle formed by $b_4, b_5, b_6, b_7$ are
not large enough to change the basically vertical orientation of the
location segments $s_{19}$, $s_{19+n}$, and $s_{20+n}$.

Now assume that there is a feasible partition, $S=S_1\stackrel{\cdot}{\cup}S_2$,
such that $\sum_{i\in S_1} a_i =\sum_{i\in S_2} a_i$.
By performing a (finite) ``wiggly'' sequence of moves, we can
move the partition segments such that
(1) any segment $s_{19+i}$ representing $a_i\in S_1$ satisfies
$y_i-y_{i-1}=a_i+O(\varepsilon^5)$ and
$|x_i-x_{i-1}|=O(\varepsilon^5)$, so that $s_{19+i}$ is pointing up;
(2) any segment $s_{19+i}$ representing $a_i\in S_2$ satisfies
$y_{i-1}-y_i=a_i+O(\varepsilon^5)$ and
$|x_i-x_{i-1}|=O(\varepsilon^5)$, so that $s_{19+i}$ is pointing down;
and (3) we end up placing $b_{20+n}$ within Euclidean distance
$O(\varepsilon^4)$ from $b_{18}$ and placing $b_{21+n}$ at distance
$\varepsilon/3+O(\varepsilon^5)$ from $b_3$ and $b_8$. Thus, extending
the spring by an appropriate wiggly motion moves the key through the
bottleneck. Now it is easy to open the joint $b_{13}$, and unfold the
whole chain.

Conversely, assume that the chain can be unfolded. As discussed above,
the sum of angles at $b_{13}$ and at $b_{17}$ has to change significantly
before the frame ceases to be rigid. Now note that
the dimensions of the bottleneck force the keyholder segment to 
be roughly vertical, i.e., 
to have slope within $O(1/\varepsilon^3)$ of vertical. 
(See Figure~\ref{fi:bottleneck}.)
Furthermore, we noted above that any feasible vertical motion of $b_{17}$
does not change the angle at $b_{17}$ by a significant amount;
it is clear that this also prevents the angle at $b_{13}$ from changing much.
Therefore, the frame remains rigid until the 
key has been removed from the lock.

\begin{figure}[htbp]
\begin{center}
  \leavevmode
  \epsfig{file=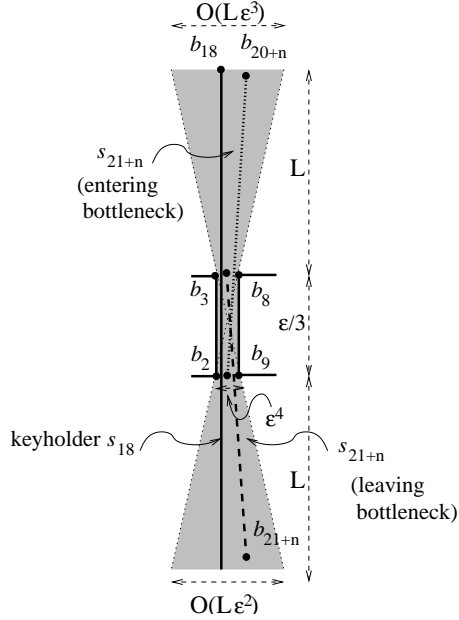,width=0.5\columnwidth}  
  \caption{All segments have to be inside of
the shaded region when moving through the bottleneck, i.e., 
must be close to being vertical. 
(Horizontal scale and size of the bottleneck are vastly exaggerated to 
allow sufficient resolution. In scale, the shaded region is basically a
vertical line.)}
  \label{fi:bottleneck}
\end{center}
\end{figure}

Now consider the positions of segments $s_{18}$ and $s_{21+n}$ 
when $b_{21+n}$ crosses the horizontal line $y=y_2$.
By the dimensions of the bottleneck, $s_{21+n}$ must have a slope within
$\Omega(1/\varepsilon^3)$ of vertical.
Furthermore, by construction of the rectangle $b_4b_5b_6b_7$, 
we are assured that $y_{18}-y_{20+n}\leq O(\varepsilon)$, i.e.,
$s_{20+n}$ cannot be significantly below $s_{18}$.
On the other hand, $b_{20+4n}$ must be below the horizontal line $y=y_3$ when
$b_{18}$ has been moved down by a vertical distance of $L+\varepsilon$.
Since no segment within the key can change its vertical slope
significantly while $s_{21+n}$ is within the bottleneck (they must all
remain wedged between $s_{18}$ and $s_{21+n}$ and
be strictly contained in the shaded region in Figure~\ref{fi:bottleneck}),
we conclude
that $y_{20+n}-y_{18}\leq O(\varepsilon)$ upon leaving the bottleneck,
i.e., $s_{20+n}$ cannot be significantly above $s_{18}$. 

Therefore, the sets 
$S_1=\{i\in 1,\ldots,n| y_i\geq y_{i-1}\}$
and
$S_2=\{i\in 1,\ldots,n| y_i< y_{i-1}\}$
upon entering the bottleneck from above and leaving it from below
must satisfy $\sum_{i\in S_1} a_i = \sum_{i\in S_2} a_i+\Theta(\varepsilon)$.
Since $\epsilon\ll 1$, this implies that there is a feasible partition.
\hfill $\Box$

\end{pf}

For the case of monotonic bend operations, the above proof can be
easily modified:

\begin{cor}
\label{cor:monotonic.degenerate}
It is NP-hard to decide if a polygonal chain $P$ with a single
vertex-to-vertex incidence can be straightened by 
monotonic partial single-joint bends.
\end{cor}

\begin{pf}
The joints in the construction shown in Figure~\ref{fi:degenerate}
that may not be changed monotonically are $b_{14},\ldots,b_{20+n}$,
the ones that are not part of the frame. 
By using small quadruple joint gadgets as in the construction
for Theorem~\ref{th:complete}, we get a chain that can be opened with
monotonic moves, if and only if it can be opened.
\hfill $\Box$

\end{pf}
 
\clearpage

\section{Algorithms}
\label{sec:algorithms}
In this section, we turn our attention to positive algorithmic
results, giving efficient algorithms for deciding if particular bend
sequences are feasible. We consider here only the case of complete
bends.

Consider an arbitrary permutation, $\sigma=(i_1,\ldots,i_n)$, of the
bends along a wire.  In order for $\sigma$ to be a foldable sequence,
it is necessary and sufficient that for each $j=1,\ldots,n$ the bend
$b_{i_j}$ is foldable.  Recall that in our notation
$P(\{i_1,\ldots,i_{j-1}\})$ denotes the 
partially bent chain after the bends at $b_{i_1},\ldots,b_{i_{j-1}}$
have been straightened.
The point $b_{i_j}$ splits
$P(\{i_1,\ldots,i_{j-1}\})$ into two subchains; let $P_0$ (resp.,
$P_{n+1}$) denote the subchain containing the endpoint $b_0$ (resp.,
$b_{n+1}$).  Now, $b_{i_j}$ is foldable if the joint at $b_{i_j}$
can be straightened without causing a collision to occur between $P_0$ and
$P_{n+1}$ at any time during the rotation about $b_{i_j}$.  We can
assume, without loss of generality, that $P_0$ is fixed and that
$P_{n+1}$ is pivoted about $b_{i_j}$.  During this bend operation at $b_{i_j}$,
each point, $u$, on $P_{n+1}$ moves along a circular arc, $A_u$,
subtending an angle $\theta_{i_j}$, centered on $b_{i_j}$.  
It is clear that in order for the bend to be feasible, none of these
arcs $A_u$ may cross the chain $P_0$, for all choices of
points $u$ on $P_{n+1}$.  

\begin{figure*}[h!btp]
\begin{center}
  \epsfig{file=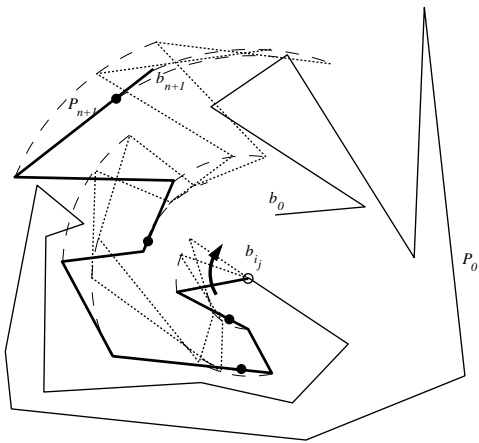,width=.52\textwidth}
  \caption{Foldability of the joint $b_{i}$: The subchain $P_{n+1}$ is shown
    with thicker lines (two dashed copies show it after different
    stages of rotation about $b_{i_j}$).  Each vertex and each
    projection point (shown as black disks) of $P_{n+1}$ moves along a
    circular arc, shown using a thin dashed arc. In this example, the
    rotation shown is {\em not} feasible, as it fails both conditions
    (1) and (2) of the lemma.}
  \label{fig:feasible-bend}
\end{center}
\end{figure*}

If the perpendicular projection of $b_{i_j}$ onto the line containing
an edge $e$ of $P_{n+1}$ lies on the edge, let $w_e\in e$ denote the
projection point.  (Each edge of $P_{n+1}$ has at most one projection
point.)  Let $U$ denote the union of the set of vertices of $P_{n+1}$
and the set of projection points on edges of $P_{n+1}$.  In the lemma
below, we observe that, in order to test feasibility of straightening
the bend $b_{i_j}$, it suffices to consider only the feasibility of
the final position of the chain $P_{n+1}$ and to test $P_0$ for
intersection with the discrete set $\mathcal{A}=\{A_u : u\in U\}$.  See
Figure~\ref{fig:feasible-bend}.

\begin{lem}
\label{lem:fold}
Joint $b_{i_j}$ is foldable if and only if (1) no arc of $\mathcal{A}$
intersects $P_0$, and (2) after the bend, no segment of $P_{n+1}$
intersects a segment of $P_0$.
\end{lem}

\begin{pf}
  If joint $b_{i_j}$ is foldable then, by definition, there can be no
  intersection of $P_{n+1}$ with $P_0$ during its rotation about
  $b_{i_j}$.  This implies conditions (1) and (2).
  
  If conditions (1) and (2) hold, then we claim that there can be no
  intersection of $P_{n+1}$ with $P_0$ during the rotation.  Consider
  a subsegment, $s$, of $P_{n+1}$ whose endpoints are consecutive
  points of $U$.  (Note that at least one endpoint of $s$ must be a
  vertex of $P_{n+1}$.)  During the rotation, it sweeps a region $R_s$
  that is bounded by two circular arcs centered at $b_{i_j}$,
  corresponding to the trajectories of its endpoints during the
  rotation, and two line segments, corresponding to the positions of
  $s$ before and after the rotation.  Here, we are using the fact that
  the distance from $b_{i_j}$ to a point $q\in s$ monotonically
  changes as a function of the position of $q$ on $s$.  (The
  projection points were introduced in order to assure this property.)
  Our claim follows from the fact that $P_0$ is a simple, connected
  chain: It cannot intersect $s$ at some intermediate stage of the
  rotation unless it intersects the boundary of the region $R_s$.
  Such an intersection is exactly what is being checked with
  conditions (1) and (2).
\hfill $\Box$
\end{pf}

\begin{lem}
\label{lem:test}
For any $S\subseteq B$, and any $b_i\not\in S$, one can decide in
$O(n\log n)$ time if joint $b_i$ is foldable for the chain~$P(S)$.
\end{lem}

\begin{pf}
  Using standard plane sweep methods for segment intersections,
  adapted to include circular arcs, we can check in $O(n\log n)$ time
  both conditions ((1) and (2)) of Lemma~\ref{lem:fold}.  Events in
  the sweep algorithm correspond to joints and to vertical points of
  tangency of circular arcs, assuming we use a vertical sweep line.
  During the sweep, we keep track of the vertical ordering of the
  segments and arcs that cross the sweep line; we check for
  intersection between any two objects that become adjacent in this
  ordering, stopping if a crossing is detected.  Since we process
  $O(n)$ events, each at a cost of $O(\log n)$, the time bound
  follows.
\hfill $\Box$
\end{pf}

\begin{rem}
  Condition (2) can be tested in linear time, by Chazelle's
  triangulation algorithm.  We suspect that condition (1) can also be
  tested in linear time.  Condition (1) involves testing for rotational
  separability of two {\em simple} chains about a fixed center point
  ($b_i$), which is essentially a polar coordinate variant of
  translational separability (which is easily tested for simple chains
  using linear-time visibility (lower envelope) calculation).  The
  issue that must be addressed for our problem, though, is the
  ``wrap-around'' effect of the rotation; we believe that this can be
  resolved and that this idea should lead to a reduction in running
  time of a factor of $\log n$.
\end{rem}

\begin{cor}
The foldability of a permutation $\sigma$ can be tested in $O(n^2\log n)$ time.
\end{cor}

We obtain improved time bounds for testing the feasibility of a
particularly important folding sequence: the identity permutation.
Many real tube-bending and wire-bending machines operate in this way,
making bends sequentially along the wire/tube.  (Such is the case for
the hydraulic tube-bending machines at Boeing's factory, where this
problem was first suggested to us.)  Of course, there are chains $P$
that can be straightened using an appropriate folding sequence but
cannot be straightened using an identity permutation folding sequence;
see Figure~\ref{fig:paperclips}(e).  However, for this special case of
identity permutations, we obtain an algorithm for determining
feasibility that runs in nearly linear time:

\begin{thm}
\label{thm:one-end-alg}
In time $O(n\log^2 n)$ one can verify if the identity permutation
($\sigma=(1,2,\ldots,n)$) is a foldable permutation for~$P$.
\end{thm}

\begin{pf}
  For notational convenience, we consider the equivalent problem of
  verifying if it is feasible, in the order $b_1,b_2,\ldots,b_n$, to
  {\em bend} the joints $b_i$ {\em from} joint angle $\pi$ {\em to}
  final angle $\theta_i$, thereby transforming a straight wire {\em
    into} the final shape $P$, rather than our convention until now of
  considering the problem of performing bend operations to straighten
  the chain~$P$.
  
  Thus, consider performing the bends in the order given by the
  identity permutation $\sigma$, and consider the moment when we are
  testing the foldability of $b_i$.  The subchain $P_{n+1}$, from
  $b_i$ to $b_{n+1}$, is a single line segment, $b_ib_{n+1}$, since
  the joints $b_{i+1},\ldots,b_{n}$ are straight at the moment.  Thus,
  verifying the foldability of bend $b_i$ amounts to testing if the
  segment $b_ib_{n+1}$ can be rotated about $b_i$ by the desired
  amount, without colliding with any other parts of the subchain $P_0$
  of $P$, from $b_0$ to $b_i$.  In other words, we must do a {\em
    wedge emptiness query} with respect to $P_0$, defined by $b_i$,
  segment $b_ib_{n+1}$, and the angle $\theta_i$.  Since $P_0$ is
  connected, emptiness can be tested by verifying that the boundary of
  the wedge does not intersect $P_0$.  (See
  Figure~\ref{fig:feasible-end-bend}.) Thus, we can perform this query
  by using (straight) ray shooting and circular-arc ray shooting in
  $P_0$; the important issue is that $P_0$ is {\em dynamically
    changing} as we proceed with more bends.

However, in order to avoid the development of potentially complex
dynamic circular-arc ray shooting data structures, we devise a simple
and efficient method that ``walks'' along portions of $P_0$, testing
for intersection with the circular arc, $\gamma$, from $b_{n+1}$ to
$b'_{n+1}$, where $b'_{n+1}$ is the location of $b_{n+1}$ {\em after}
the bend at $b_i$ has been performed.

\begin{figure}[htbp]
\begin{center}
  \epsfig{file=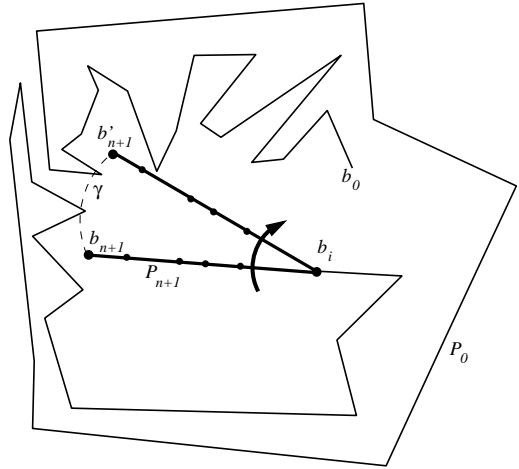,scale=0.34}   
  \caption{Testing the foldability of the joint $b_{i}$.   This example is intended to illustrate a generic step in the algorithm; for this particular chain, note that
    it is not feasible to make the bends $b_1,\ldots,b_{i-1}$ to get
    to the state shown.}
  \label{fig:feasible-end-bend}
\end{center}
\end{figure}

In particular, we keep track of a ``painted'' portion of $P_0$, which
corresponds to the subset of $P_0$ that has been ``walked over.''  We
consider the chain $P_0$ to be a degenerate simple polygon, having two
{\em sides} which form a counterclockwise loop around $P_0$.  We
consider the case in which the bend at $b_i$ is a rotation of the
segment $b_ib_{n+1}$ {\em clockwise} to the segment $b_ib'_{n+1}$; the
case of a counterclockwise bend at $b_i$ is handled similarly.  When
we perform a bend at $b_i$, we walk (counterclockwise) along the {\em
  unpainted} portions of $P_0$, between two points, $a$ and $a'$, on
the boundary of $P_0$, where $a$ and $a'$ are defined according to
cases that depend on the outcomes of two ray-shooting queries:

\begin{figure}[h!tbp]
\begin{center}
  \epsfig{file=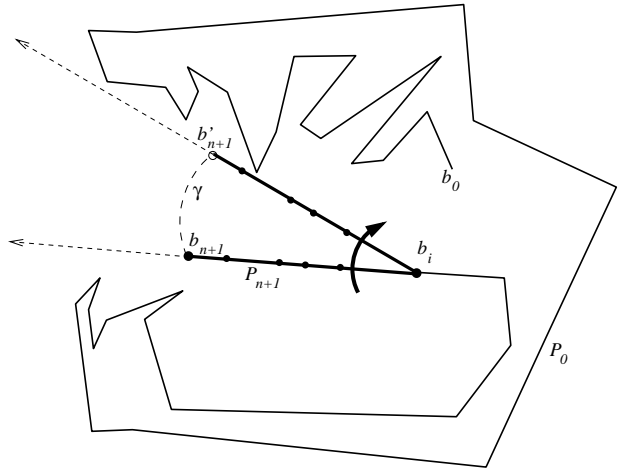,scale=0.34}
  \caption{Case (a): 
Both of the rays $b_ib_{n+1}$ and
$b_ib'_{n+1}$ miss $P_0$ and go off to infinity.}
  \label{fig:case-a}
\end{center}
\end{figure}

\begin{description}
\item[(a)] If both of the rays $\overrightarrow{b_ib_{n+1}}$ and
  $\overrightarrow{b_ib'_{n+1}}$ miss $P_0$ (and go off to infinity),
  then there is nothing more to check: the rotation at $b_i$ can be
  done without interference with $P_0$, since $P_0$ is a (connected)
  polygonal chain lying in the complement of the wedge defined by
  $\overrightarrow{b_ib_{n+1}}$ and $\overrightarrow{b_ib'_{n+1}}$.
  See Figure~\ref{fig:case-a}.
  
\begin{figure}[h!tbp]
\begin{center}
  \epsfig{file=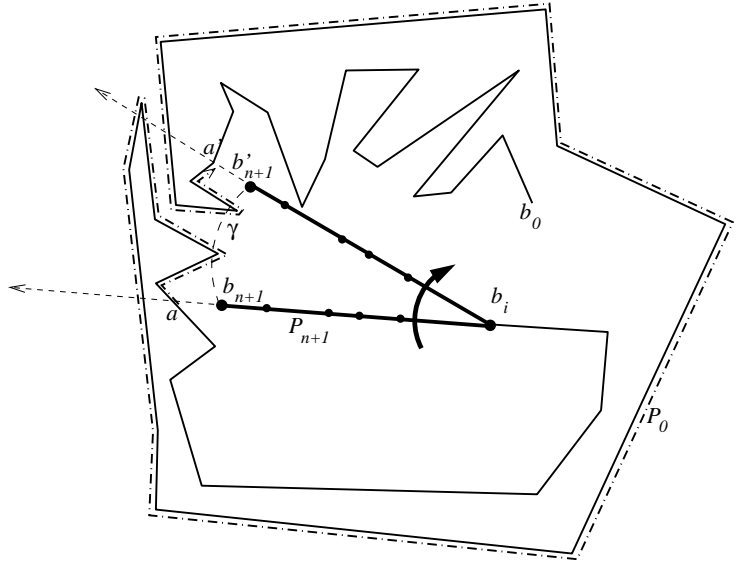,scale=0.4}
  \caption{Case (b): 
Both of the rays $b_ib_{n+1}$ and
  $b_ib'_{n+1}$ hit $P_0$.  The walk extends from $a$ to $a'$ over the highlighted
portion of $P_0$, painting any previously unpainted portion of it.}
  \label{fig:case-b}
\end{center}
\end{figure}

\item[(b)] If both of the rays $\overrightarrow{b_ib_{n+1}}$ and
  $\overrightarrow{b_ib'_{n+1}}$ hit $P_0$, then we let $a$ and $a'$
  (respectively) be the points on the boundary of $P_0$ where they first hit
  $P_0$.  See Figure~\ref{fig:case-b}.
  
\begin{figure}[h!tbp]
\begin{center}
  \epsfig{file=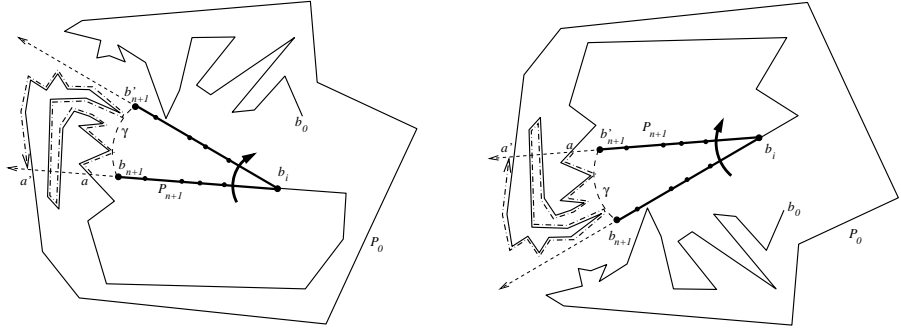,width=.98\textwidth}
  \caption{Case (c): 
Exactly one of the rays $b_ib_{n+1}$ and
  $b_ib'_{n+1}$ hits $P_0$: $b_ib_{n+1}$ hits $P_0$ (left) or $b_ib'_{n+1}$
hits $P_0$ (right). The walk extends from $a$ to $a'$ over the highlighted
portion of $P_0$, painting any previously unpainted portion of it.}
  \label{fig:case-c}
\end{center}
\end{figure}

\item[(c)] If exactly one of the rays $\overrightarrow{b_ib_{n+1}}$
  and $\overrightarrow{b_ib'_{n+1}}$ hits $P_0$ while the other misses
  $P_0$ (and goes off to infinity), then we define $a$ and $a'$ as
  follows.  Assume that the ray $\overrightarrow{b_ib_{n+1}}$ hits
  $P_0$ (and the ray $\overrightarrow{b_ib'_{n+1}}$ misses $P_0$); the
  other case is handled similarly (see Figure~\ref{fig:case-c}, right).  
  Then, we define $a$ to be the point on
  the boundary of $P_0$ where the ray $\overrightarrow{b_ib_{n+1}}$
  hits $P_0$, and we define $a'$ to be the point on the boundary of
  $P_0$ where a ray from infinity in the direction
  $\overrightarrow{b_{n+1}b_i}$ (towards $b_i$) hits $P_0$.  See
  Figure~\ref{fig:case-c}, left.
\end{description}

During the walk from $a$ to $a'$ along the boundary of $P_0$, we test
each segment for intersection with the circular arc $\gamma$ in time
$O(1)$.  Whenever we reach a portion of the boundary that is already
painted, we skip over that portion, going immediately to its end. 
Already painted portions have endpoints that were
determined by rays in previous steps of the painting procedure.
Since there are only a total of $O(n)$ rays (one per edge of $P$),
this implies only $O(n)$ endpoints of painted portions.  As
we walk, we mark the corresponding portions over which we walk as
``painted.''
Since, by continuity, it is easy to see that the painted portion of
any one segment of $P_0$ is connected, we know that we must encounter
at least one vertex of $P_0$ between the time that the walk leaves a
painted portion and the time that the walk enters the next painted
portion.  Thus, during a walk, we charge the tests that we do for
intersection with $\gamma$ off to the vertices that are being painted.
The remainder of the justification of the algorithm is based on
two simple claims:

\begin{claim}
There is no need to walk back over a painted portion in order to
check for intersections with an arc $\gamma$ at some later stage.
\end{claim}

\begin{pf}
The fact that we need not walk over a painted portion testing again
for intersections with $\gamma$ follows from the fact that with each
bend in the sequence, the length of the segment $b_ib_{n+1}$ that we
are rotating goes {\em down} by the length of the last link.  Thus, 
if the motion of the tip, $b_{n+1}$, sweeps an arc
$\gamma$ that does not reach a portion $\mu$ of the boundary of $P_0$
when the link $b_ib_{n+1}$ is {\em straight}, it cannot later be that
a link $b_jb_{n+1}$ ($j>i$) can permit the tip $b_{n+1}$ to reach the
same portion $\mu$ when pivoting is done about $b_j$; this is a
consequence of the triangle inequality.
\hfill $\Box$
\end{pf}

\begin{claim}
  In testing for intersection with $\gamma$, we check enough of the
  chain $P_0$: if any part of it intersects $\gamma$, then it must lie
  on the portion between $a$ and $a'$ over which we walk.
\end{claim}

\begin{pf}
  In case (a), there is nothing to check.  In case (b), the closed
  Jordan curve from $b_i$ to $a$ (along a straight segment), then
  along the boundary of the simple polygon $P_0$ to $a'$, then back to
  $b_i$ (along a straight segment) forms the boundary of a region
  whose only intersection with $P_0$ is along the shared boundary from
  $a$ to $a'$; thus, if $\gamma$ lies within this region (i.e., does
  not intersect the boundary of $P_0$ from $a$ to $a'$), then $\gamma$
  does not intersect any other portion of $P_0$.  In case (c) we argue
  similarly, but we use the Jordan region defined by the segment from
  $b_i$ to $a$, the boundary of $P_0$ from $a$ to $a'$, the ray from
  $a'$ to infinity (in the direction of
  $\overrightarrow{b_ib_{n+1}}$), then the reverse of the ray
  $\overrightarrow{b_ib'_{n+1}}$ back to~$b_i$.
\hfill $\Box$
\end{pf}

The total time for walking along the chain $P_0$ can be charged off to
the vertices of $P$, resulting in time $O(n)$ for tests of
intersection with arcs $\gamma$, exclusive of the ray shooting time.
The final time bound is then dominated by the time to perform $n$
straight ray shooting queries in a dynamic data structure for the
changing polygonal chain $P_0$; these ray shooting queries are
utilized both in testing for intersection with the segment
$b_ib'_{n+1}$ and in determining the points $a$ and $a'$ that define
the walk.  These ray-shooting queries and updates are done in time
$O(\log^2 n)$ each, using existing techniques (\cite{gt-drssp-97}),
leading to the claimed overall time bound.
\hfill $\Box$
\end{pf}

\begin{figure}[bhtp]
\begin{center}
  \epsfig{file=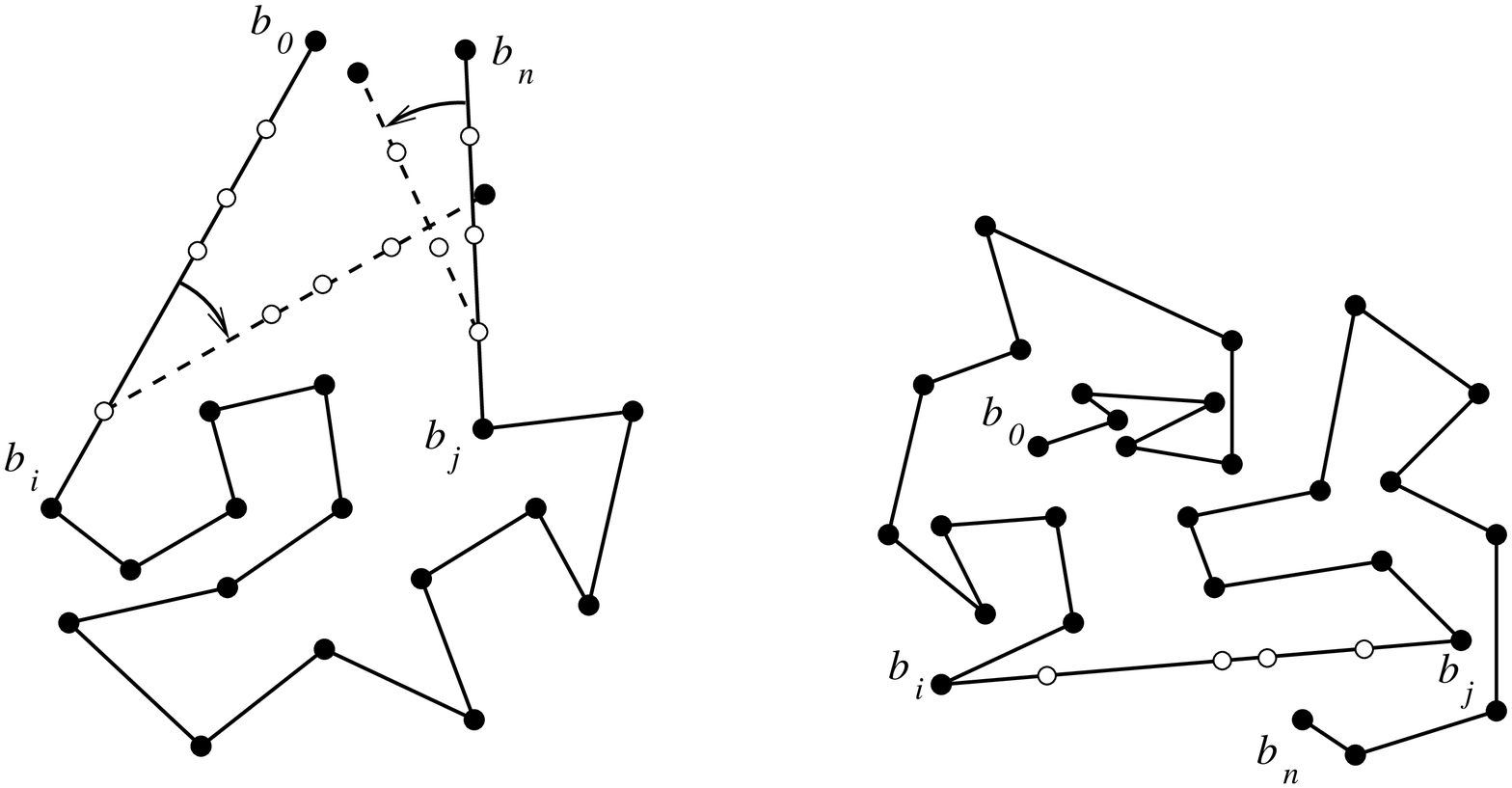,width=.82\textwidth}
  \caption{An intermediate state $(i,j)$ in the
    bending of an outwards (left) and an inwards (right) folding
    sequence.  For the outwards folding sequence on the left, the next
    bend is either $b_{i-1}$ or $b_{j+1}$; the new positions of the
    chain are shown dashed.  For the inwards folding sequence on the
    right, the next bend is either $b_{i+1}$ or $b_{j-1}$.}
  \label{fig:inwards-outwards}
\end{center}
\end{figure}

Next, we turn to two other important classes of permutations.  Again,
for notational convenience, we consider the problem of verifying if it
is feasible, in the order given by the permutation, to {\em bend} the
joints $b_i$ {\em from} joint angle $\pi$ {\em to} final angle
$\theta_i$, thereby transforming a straight wire {\em into} the final
shape $P$.  We say that a permutation is an {\em outwards folding
  sequence} (resp., {\em inwards folding sequence}) if at any stage of
the folding, the set of bends that have been completed, and therefore
are {\em not} straight, is a subinterval, $b_i,b_{i+1},\ldots,b_j$
(resp., a pair of intervals $b_1,b_2,\ldots,b_i$ and
$b_j,b_{j+1},\ldots,b_n$); thus, the next bend to be performed is
either $b_{i-1}$ or $b_{j+1}$ (resp., $b_{i+1}$ or $b_{j-1}$).
Inwards and outwards folding sequences are a subclass of permutations
that model a constraint imposed by some forming machines.  See
Figure~\ref{fig:inwards-outwards}.  The identity permutation is a
folding sequence that is a special case of both an inwards and an
outwards folding sequence.

We show that one can efficiently search for a folding sequence that is
inwards or outwards.  Our algorithms are based on dynamic programming.

\bigskip
First, consider the case of outwards folding sequences.  We keep track
of the {\em state} as the pair $(i,j)$ representing the interval of
bends ($b_i,b_{i+1},\ldots,b_j$) already completed.  We construct a
graph $\mathcal{G}$ whose $O(n^2)$ nodes are the states $(i,j)$ (with
$1\leq i<j\leq n$) and whose edges link states that correspond to the
action of completing a bend at $b_{i-1}$ or $b_{j+1}$ (if the starting
state is $(i,j)$).  Thus, each node has constant degree.  Our goal is
to determine if there is a path in this graph from some $(i,i)$, for
$i\in\{1,2,\ldots,n\}$ to $(1,n)$.  
We augment this graph with a special node $\nu_0$, linked to
each node $(i,i)$.  Then, our problem is readily solved in $O(n^2)$ time once
we have the graph constructed, since it is simply searching for a path
from node $\nu_0$ to node $(1,n)$. (Alternatively, we can construct the graph as we
search the graph for a path.)  In order to construct the graph, we
need to test whether bend $b_{i-1}$ or $b_{j+1}$ can be performed
without intersecting the folded chain, $P'$, linking $b_{i-1}$ to
$b_{j+1}$.  This is done in a manner very similar to that we described
above for the case of identity permutations: we perform ray-shooting
queries in time $O(\log^2 n)$ and then use a ``painting'' procedure to
keep track of the states of $2n-2$ ``walks'' that determine
circular-arc ray shooting queries.  In particular, there is a separate
painting procedure corresponding to each of the $n-1$ choices of $i$
and to each of the $n-1$ choices of $j$.  For example, for a fixed
choice of $i$, the painting procedure will consider each of the
possible bends $b_{i+1},\ldots,b_n$ in order, allowing us to amortize
the cost of checking for intersections with the circular arc $\gamma$
associated with each bend.  In total, the cost of the walks is
$O(n^2)$, while there may also be $O(n^2)$ ray shooting queries (in a
dynamically changing polygon).  Thus, the total cost is dominated by
the ray shooting queries, giving an overall time bound of $O(n^2\log
n)$.

For the case of an inwards folding sequence, we build a similar state
graph and search it.  However, the cost of testing if a bend is
feasible is somewhat higher, as we do not have an especially efficient
procedure for testing the foldability of a polygonal chain.  (Our
painting procedures exploit the fact that the link being folded is
straight.)  Thus, we apply the relatively naive method of testing
feasibility given in Lemma~\ref{lem:test}, at a cost of $O(n\log n)$
per test (which potentially improves to $O(n)$ time, if our conjecture
mentioned in the remark after the Lemma is true).  Thus, the overall
cost of the algorithm is dominated by the $O(n^2)$ feasibility tests,
at a total cost of $O(n^3\log n)$.  In summary, we have:

\begin{thm}
\label{thm:two-ends-alg}
In time $O(n^2\log^2 n)$ one can determine if there is an outwards
folding sequence; in time $O(n^3\log n)$ one can determine if there
is an inwards folding sequence.
\end{thm}

\clearpage
\section{Conclusion}

We conclude with some open problems that are suggested by our work:
\begin{description}
\item[(1)] Is the bend sequencing problem for wire folding {\em strongly}
  NP-complete, or is there a pseudo-polynomial-time 
  algorithm?  If not
  in wire bending, is it strongly NP-complete for the 3-dimensional
  sheet metal folding problem?
\item[(2)] Is it NP-hard to decide if a polygonal chain in three
  dimensions can be straightened?  In \cite{bddlloorstw-lupc3d-99} 
  simple examples of locked chains in three dimensions are shown; 
  can
  these be extended to a hardness proof for the decision problem?
\item[(3)] In practice, in order to make a bend using a punch and die
  on a press brake, it is necessary to consider accessibility
  constraints.  For each bend operation, the die is placed on one side
  of the material, while the punch is placed on the other side.  The
  bend is formed by pushing the punch into the die (which has a
  matching shape), with the material in between.  (See
  Wang~\cite{w-mddsmp-97}.)  In the simplest model of this operation
  on a wire, we can consider the punch and the die to be oppositely
  directed rays that form a bend by coming together (from opposite
  sides of the wire) so that their apices meet at the bend point.  The
  accessibility constraint in this simple model is that the rays
  representing the punch and die must be disjoint from the wire
  structure both at the initial placement of these ``tools'' and
  during the bend operation itself.
\item[(4)] Can the foldability of a permutation be decided in
  subquadratic time for wire bending?  This would be possible if one
  had a dynamic data structure that will permit efficient (sublinear)
  queries for the foldability of a vertex.
\end{description}

\subsection*{Acknowledgments}
We thank an anonymous referee for pointing out some errors in an
earlier draft and for many valuable suggestions to improve the
presentation.  We thank Steve~Skiena for valuable input in the early
stages of this research and, in particular, for contributing ideas to
the hardness proof of Theorem~\ref{th:complete}.  We also thank Erik
Demaine for discussions leading to the current version of
Theorem~\ref{th:simple}.  Estie~Arkin acknowledges support from the
National Science Foundation (CCR-9732221,CCR-0098172) and HRL
Laboratories. S\'andor~Fekete acknowledges support from the National
Science Foundation (ECSE-8857642,CCR-9204585) during his time at Stony
Brook (1992-93), when this research was initiated.  Joe~Mitchell
acknowledges support from HRL Laboratories, the National Science
Foundation (CCR-9732221,CCR-0098172), NASA Ames Research Center,
Northrop-Grumman Corporation, Sandia National Labs, Seagull
Technology, and Sun Microsystems.


\bibliography{refs}

\end{document}